\documentclass[onecolumn,preprintnumbers,amsmath,amssymb]{revtex4}
\usepackage{epsfig}
\usepackage{amsmath}
\begin{document}
\title{Theory of Electronic Relaxation in solution with ultra-short sink of different shapes: An exact analytical solution}
\author{Swati Mudra* and Aniruddha Chakraborty \\
School of Basic Sciences, Indian Institute of Technology Mandi,\\
Kamand, Himachal Pradesh, 175005, India}
\date{\today}
\begin{abstract} 
\noindent We propose a  very simple one dimensional analytically solvable model for understanding the problem of electronic relaxation of molecules in solution. This problem is modeled by a particle diffusing under the influence of parabolic potential in presence of a sink of ultra-short width. The diffusive motion is described by the Smoluchowski equation and shape of the sink is represented by  1) ultra-short Gaussian, 2) ultra-short exponential and 3) ultra-short rectangular function at arbitrary position. Rate constants are found to be sensitive to the shape of the sink function, even though the width of the sink is too small. This model is of considerable importance as a realistic model in comparison with the point sink model for understanding the problem of electronic relaxation of a molecule in solution.  
\end{abstract}

\maketitle

\section{Introduction}
\noindent Electronic relaxation of a molecule immersed in a polar solvent is an interesting research problem and there are many attempts to understand this problem \cite{Harris,Ben,Kls1,Ani1,Bagchi,lippert,SK,Ani2,Robin,Amb,Kls3}. A molecule immersed in a polar solvent can be put on an electronically excited state by using radiation of appropriate frequency. The molecular configuration executes a random walk on excited state potential energy surface as a result of interaction between the molecule and solvent. As the molecular configuration changes randomly on that potential energy surface, it may undergo non-radiative decay from certain regions of that surface. It may also undergo radiative decay from any configuration from that surface with equal probability. From the theoretical point of view, the problem is to calculate the survival probability of that molecule on that electronically excited state potential energy surface. In the following we use  $P_{e} (t)$, to denote the survival probability of the molecule on that electronically excited state potential curve. In the following we assumes one dimensional model for the molecular configuration and the relevant coordinate is denoted by $x$ - this is generally the assumption in all existing models in literature \cite{Kls1,Ani1,Bagchi,SK,Fleming}. Therefore, in the discussion below, we will use the variable $x$ to denote a particular configuration of the molecule and de-excitation of the molecule as the absorption. Therefore, the probability $P(x,t)$ that the molecule can be found in the configuration $x$ at the time $t$ follows the following Smoluchowski equation with a sink term.
\begin{equation}
\frac{\partial P(x,t)}{\partial t} = [L - k_{r} - k_{o} S(x)] P(x,t),
\end{equation}
\noindent where $S(x)$ is the sink function which depends on the molecular configuration, $k_{0}$ is the rate constant for non-radiative decay process, $k_{r}$ is the rate constant for radiative decay process and the operator $L$ is given by
\begin{equation}
    L = D\frac{\partial^2}{\partial x^2} + \frac{D}{k_{b}T}\frac{\partial}{\partial x}\left[\frac{\partial V(x)}{\partial x}\right].
\end{equation}
\noindent In the above $V(x)$ is the potential which is a function of molecular configuration and is assumed to be harmonic for electronically excited state potential energy curve - which is a very standard assumption.
\begin{equation}
V(x) = \frac{Bx^2}{2}.
\end{equation}
\noindent In the above $D$ is a diffusion constant. Initially the molecule is in electronically ground state and is immersed in a solvent at a finite temperature $T$ and its distribution over the configurational coordinate can be assumed to be random. From this state the molecule undergoes Franck-Condon excitation (molecular configuration does not change on excitation) to the excited electronic state potential energy curve. As a result $x_0$ the initial configuration of the molecule on the excited state is assumed to be random. There are only few model problems of this kind for which exact solutions have been found  \cite{Kls1,Ani1,Bagchi,SK,Ani2,Kls3}. The Oster-Nishijima model \cite{Oster} assumes that the molecular configuration changes freely [the corresponding potential is zero {\it i.e.}, $V(x) = 0$] in the region where $0 < x < a $ but when it goes out of the region, it decays with unit probability. The pinhole sink model \cite{Bagchi,SSZ} has a hole in the potential energy curve whose width is tending to zero. On reaching this configuration the molecule would decay with unit probability. Mathematically this sink function is taken to be a Dirac delta function of infinite strength \cite{Bagchi,Fleming}. Exact analytical solution in time domain can be obtained in the case where $V(x)$ is parabolic and this pinhole sink is at the origin. Cases where the sink is not placed at the origin and strength of that sink is finite are very interesting and can be solved only in Laplace domain \cite{Kls1}. In that case, exact analytical expressions for different rate constants for parabolic potential is derived by Sebastian \cite{Kls1}. Cases where the sink is assumed to be a Gaussian function for motion under parabolic potential is studied numerically by Bagchi {\it et. al.,} \cite{bagchi1}. All the  analytically solvable model assumed `point' sink, but sink function is expected to have a finite width \cite{Alok}. Recently we proposed one analytically solvable model where the sink function is non-zero for a narrow width \cite{SM}. In this paper we give a general procedure for finding the exact analytical solution for the same problem with  1) ultra-short gaussian sink, 2) ultra-short exponential sink  and 3) ultra-short rectangular sink, for parabolic potential.
\noindent The Eq. (1) can be written as \cite{Kls1}
\begin{equation}
\frac{\partial P(x,t)}{\partial t} = D\frac{\partial^2 P(x,t)}{\partial x^2} + \frac{DB}{k_{b}T} \frac{\partial}{\partial x} x P(x,t) - k_{o} S(x) P(x,t) - k_{r} P(x,t).
\end{equation}
\noindent Laplace transformation of ${\tilde P}(x,s)$ of $P(x,t)$  can be defined by 
\begin{equation}
\tilde P(x,s)= \int^\infty_0 P(x,t) e^{-st} dt.
\end{equation}
\noindent So Eq.(3) can be written in Laplace domain as given by
\begin{equation}
\left[s {\tilde P}(x,s)-D\frac{\partial^2{\tilde P}(x,s)}{\partial x^2} - \frac{DB}{k_{b}T}\frac{\partial}{\partial x} \left(x {\tilde P}(x,s)\right) + k_{o} S(x){\tilde P}(x,s)+{k_{r}}{\tilde P}(x,s)\right] =  P(x,0).
\end{equation}
\noindent In the following, we give a method for finding the exact solution of the problem in case, 1) ultra-short Gaussian sink, 2) ultra-short exponential sink and 3) ultra-short rectangular sink.

\section{Exact Analytical results}
\subsection{Ultra-short Gaussian Sink}

\noindent Here we give an exact analytical solution of Eq. (6) in the case where sink is represented by a Gaussian function of ultra-short width.  Gaussian sink function is generally used for the problem of electronic relaxation of molecule in solution \cite{bagchi1} as well as few related problems including understanding protein-DNA interaction \cite{Hansen}. Now we start our calculation by replacing $S(x)$ of Eq. (6), by a truncated Gaussian function {\it i.e.}, given by
\begin{equation}
S(x) = \frac{\sqrt{{\alpha_G}}}{\sqrt{\pi} Erf(\sqrt{{\alpha_G}} \epsilon)} e^{- {\alpha_G} (x - x_c)^2} f(x).  \end{equation}    
\noindent In the above $S(x)$ is non zero within an ultra-short range of $x$-coordinate, {\it i.e.,} from  $x_c - \epsilon$ to $x_c + \epsilon$, where $\epsilon$ is a  positive number very close to zero. Outside this ultra-short range the value of $S(x)$ is zero. The above sink function is normalized appropriately, therefore $\int_{-\infty}^{\infty} S(x) dx = 1$. The additional function $f(x)$ is introduced in the definition of sink function just to ensure that $S(x)$ is definitely zero outside this range $x_c - \epsilon$ and $x_c + \epsilon$ (therefore $f(x)$ is a function which is  $1$ within this ultra-short range and $f(x)$ is zero outside this range). Therefore in Eq.(6), the term $k_{o} S(x){\tilde P}(x,s)$ may be replaced by $k_{0} e^{- {\alpha_G} (x - x_c)^2}{\tilde P}(x_c,s) f(x)$ and therefore Eq. (6) becomes
\begin{equation}
\left[s {\tilde P}(x,s)- D\frac{\partial^2{\tilde P}(x,s)}{\partial x^2} - \frac{DB}{k_{b}T}\frac{\partial{\tilde P}(x,s)}{\partial x} x + \frac{k_{0}\sqrt{{\alpha_G}}}{\sqrt{\pi} Erf(\sqrt{{\alpha_G}} \epsilon)} e^{- {\alpha_G} (x - x_c)^2} {\tilde P}(x_c,s) f(x)+k_{r}{\tilde P}(x,s)\right] =  P(x,0).
\end{equation}
\noindent The solution of Eq. (8) may be expressed using Green's function as given below
\begin{equation}
\tilde P(x,s)= \int^\infty_{-\infty} dx_{0}G(x,s+k_{r}|x_0)P(x_0,0) - \frac{k_0 \sqrt{{\alpha_G}}}{\sqrt{\pi} Erf(\sqrt{{\alpha_G}} \epsilon)}  {\tilde P}(x_c,s)\int^{\infty}_{-\infty} dx_{0}G(x,s+k_{r}|x_0)e^{- {\alpha_G} (x_0 - x_c)^2} f(x_0),
\end{equation}
\noindent where $G(x,s|x_0)$ is the Green's function appropriate for parabolic potential in the absence of any sink. This equation can be simplified further by using the properties of $f(x)$ as shown below
\begin{equation}
\tilde P(x,s)= \int^\infty_{-\infty} dx_{0}G(x,s+k_{r}|x_0)P(x_0,0) - \frac{k_0 \sqrt{{\alpha_G}}}{\sqrt{\pi} Erf(\sqrt{{\alpha_G}} \epsilon)} {\tilde P}(x_c,s)\int^{x_c+\epsilon}_{x_c -\epsilon} dx_{0}G(x,s+k_{r}|x_0) e^{- {\alpha_G} (x_0 - x_c)^2}.
\end{equation}
\noindent In the last term on the R.H.S. of the above equation, integration is over $x_0$ from  $x_c - \epsilon$ to $x_c + \epsilon$, as this range is actually ultra-short we can easily replace  $G(x,s+k_{r}|x_0)$ by $G(x,s+k_{r}|x_c)$. Therefore Eq. (10) can be expressed as given by
\begin{equation}
\tilde P(x,s)= \int^\infty_{-\infty} dx_{0}G(x,s+k_{r}|x_0)P(x_0,0) - \frac{k_{0}\sqrt{{\alpha_G}}}{\sqrt{\pi} Erf(\sqrt{{\alpha_G}} \epsilon)}  {\tilde P}(x_c,s)G(x,s+k_{r}|x_c)\int^{x_c+\epsilon}_{x_c -\epsilon} dx_{0}e^{- {\alpha_G} (x_0 - x_c)^2}.
\end{equation}
\noindent Now we assume the case where ${\alpha_G}$ is very large, such that $e^{- {\alpha_G} (x_0 - x_c)^2}$ is practically zero, outside this ultra-short range  $x_c - \epsilon$ to $x_c + \epsilon$. Therefore, we can safely assume that the integrand {\it i.e.}, $e^{- {\alpha_G} (x_0 - x_c)^2}$ does not change within the range of integration. Therefore after we perform the integration, we get the following equation 
\begin{equation}
\tilde P(x,s)= \int^\infty_{-\infty} dx_{0}G(x,s+k_{r}|x_0)P(x_0,0) -   \frac{ 2 \epsilon k_{0} \sqrt{{\alpha_G}}}{\sqrt{\pi} Erf(\sqrt{{\alpha_G}} \epsilon)}  {\tilde P}(x_c,s)G(x,s+k_{r}|x_c).
\end{equation}
\noindent Now there are two unknowns $\tilde P(x,s)$ and $\tilde P(x_c,s)$ in the above equation, therefore we put $x=x_c$ in the above equation and get the following equation with one unknown {\it i.e.}, $\tilde P(x_c,s)$.
\begin{equation}
\tilde P(x_c,s)= \int^\infty_{-\infty} dx_{0}G(x_c,s+k_{r}|x_0)P(x_0,0) -  \frac{ 2 \epsilon k_{0}\sqrt{{\alpha_G}}}{\sqrt{\pi} Erf(\sqrt{{\alpha_G}} \epsilon)}{\tilde P}(x_c,s)G(x_c,s+k_{r}|x_c).
\end{equation}
\noindent Now we solve Eq. (14) to find the solution of $\tilde P(x_c,s)$ 
\begin{equation}
\tilde P(x_c,s)= \frac{\int^\infty_{-\infty} dx_{0}G(x_c,s+k_{r}|x_0)P(x_0,0)}{1+ \frac{ 2 \epsilon k_{0}\sqrt{{\alpha_G}}}{\sqrt{\pi} Erf(\sqrt{{\alpha_G}} \epsilon)} G(x_c,s+k_{r}|x_c)}.
\end{equation}
\noindent Now this solution, when substituted back into Eq. (13) gives
\begin{equation}
\tilde P(x,s)= \int^\infty_{-\infty} dx_{0} \left[G(x,s+k_{r}|x_0)- \frac{ \frac{ 2 \epsilon k_{0}\sqrt{{\alpha_G}}}{\sqrt{\pi} Erf(\sqrt{{\alpha_G}} \epsilon)} G(x,s+k_{r}|x_c)G(x_c,s+k_{r}|x_0)}{1+ \frac{ 2 \epsilon k_{0}\sqrt{{\alpha_G}}}{\sqrt{\pi} Erf(\sqrt{{\alpha_G}} \epsilon)} G(x_c,s+k_{r}|x_c)}\right] P(x_0,0).
\end{equation}
\noindent The above equation gives an analytical formula for $\tilde P(x,s)$. Now we are interested to calculate survival probability in Laplace domain and that is given by $P_{e}(s)$. The analytical expression of survival probability can be derived using Eq. (16) as shown below 
\begin{equation}
P_e(s) = \int^\infty_{-\infty} dx {\tilde P}(x,s).
\end{equation} 
\noindent Using the property of Green's function we get the following condition
\begin{equation}
\int^\infty_{-\infty} dx_{0} (G(x,s|x_0) = 1/s.
\end{equation}
\noindent Using the above equation we derive the following expression for $P_{e}(s)$
\begin{equation}
P_e(s)=\frac{1}{s+k_{r}}\left[1-\left(1+ \frac{ 2 \epsilon k_{0}\sqrt{{\alpha_G}}}{\sqrt{\pi} Erf(\sqrt{{\alpha_G}} \epsilon)} G(x_c,s+ k_{r}|x_c) \right)^{-1} \frac{ 2 \epsilon k_{0}\sqrt{{\alpha_G}}}{\sqrt{\pi} Erf(\sqrt{{\alpha_G}} \epsilon)} \int^\infty_{-\infty} dx_0 G (x_c,s+k_{r}|x_0)P(x_0,0)\right].
\end{equation}
\noindent Now we calculate average and long time rate constants using the analytical expression of $P_e(s)$. The average rate constant is given by $k^{-1}_I =P_e(0)$ and the long time rate constant $k_L$ = negative of the pole of $P_e(s),$ which is closest to the origin. From Eq. (19), we obtain the following expression of average rate constant
\begin{equation}
k^{-1}_I =\frac{1}{k_{r}}\left[1- \left(1+ \frac{ 2 \epsilon k_{0}\sqrt{{\alpha_G}}}{\sqrt{\pi} Erf(\sqrt{{\alpha_G}} \epsilon)} G(x_c,k_{r}|x_c) \right)^{-1} \frac{ 2 \epsilon k_{0}\sqrt{{\alpha_G}}}{\sqrt{\pi} Erf(\sqrt{{\alpha_G}} \epsilon)} \int^\infty_{-\infty} dx_0 G(x_c,k_{r}|x_0)P(x_0,0)\right].
\end{equation}
\noindent Here we can see that $k_I$ depends on the initial probability distribution $P(x,0)$ and $k_L = - $ pole of $[ 1+\frac{ 2 \epsilon k_{0}\sqrt{{\alpha_G}}}{\sqrt{\pi} Erf(\sqrt{{\alpha_G}} \epsilon)} G(x_c, s+k_r|x_c)][s+k_r]^{-1}$, the one which is closest to the origin, on the negative $s$ - axis, and is independent of the initial distribution $P(x_0,0)$. We can find $G(x,s;x_0)$ by using the following equation for harmonic potential \cite{Kls2}
\begin{equation}
\left(s - {\cal L}\right) G(x,s;x_0)= \delta (x - x_0). 
\end{equation}
\noindent Using standard method \cite{Hilbert} one can get
\begin{equation}
G(x,s;x_0)=F(z,s;z_0)/(s+k_r).
\end{equation}
\noindent with
\begin{equation}
F(z,s;z_0)= D_\nu(-z_<)D_\nu(z_>)e^{(z_0^2-z^2)/4}\Gamma(1-\nu)[B/(2 \pi D)]^{1/2}. 
\end{equation}
\noindent In the above, $z$ defined by $z = x(D/B)^{1/2}$  and $z_j = x_j(D/B)^{1/2}$, $\nu  = -s/B$ and $\Gamma(\nu)$ is the gamma function. Also, $z_{<}= min(z, z_0)$ and $z_{>}= max(z, z_0)$. $D_{\nu}$ represents parabolic cylinder functions. To understand the behavior of $k_I$ and $k_L$, we assume the initial distribution $P^0_e(x_0)$ is represented by a Dirac delta function located at $x_0$. Then, we get 
\begin{equation}
{k_I}^{-1}= (k_r)^{-1}\left(1 - \frac{\frac{ 2 \epsilon k_{0}\sqrt{{\alpha_G}}}{\sqrt{\pi} Erf(\sqrt{{\alpha_G}} \epsilon)} F(z_s,k_r|z_0)}{k_r+ \frac{ 2 \epsilon k_{0}\sqrt{{\alpha_G}}}{\sqrt{\pi} Erf(\sqrt{{\alpha_G}} \epsilon)} F(z_s,k_r|z_s)} \right).
\end{equation}
Again
\begin{equation}
k_L= k_r - [ values \; of \; s \; for \; which \;\; s+ \frac{ 2 \epsilon k_{0}\sqrt{{\alpha_G}}}{\sqrt{\pi} Erf(\sqrt{{\alpha_G}} \epsilon)} F(z_s,s|z_s)=0].
\end{equation}
\noindent We should mention that $k_I$ is dependent on the initial position $x_0$ and the value of $k_r$ but $k_L$ is independent of the initial position. For further simplification we assume $k_r\rightarrow$ 0, in this limit the solution we get, that is still valid even when $k_r$ is nonzero. By using the properties of parabolic cylindrical functions $D_v{(z)}$ \cite{Erdelyi}, we find that when $k_r\rightarrow 0, F{(z_s,k_r|z_0)}$ and $F{(z_s,k_r|z_s)}\rightarrow exp(-z_s^2/2){[B/(2\pi D)]}^\frac{1}{2}$ so that
\begin{equation}
 \frac{ 2 \epsilon k_{0}\sqrt{{\alpha_G}}}{\sqrt{\pi} Erf(\sqrt{{\alpha_G}} \epsilon)} F{(z_s,k_r|z_0)}/[k_r+  \frac{ 2 \epsilon k_{0}\sqrt{{\alpha_G}}}{\sqrt{\pi} Erf(\sqrt{{\alpha_G}} \epsilon)} F{(z_s,k_r|z_s)}]\rightarrow 1.
\end{equation}
\noindent Hence, we get the following expression for average rate constant
\begin{equation}
k_I^{-1}=-{\frac{\partial}{\partial k_r}\left[\frac{\frac{ 2 \epsilon k_{0}\sqrt{{\alpha_G}}}{\sqrt{\pi} Erf(\sqrt{{\alpha_G}} \epsilon)} F(z_s,k_r|z_0)}{k_r + \frac{ 2 \epsilon k_{0}\sqrt{{\alpha_G}}}{\sqrt{\pi} Erf(\sqrt{{\alpha_G}} \epsilon)} F(z_s,k_r|z_s)}\right]}_{(k_r) \rightarrow 0}.
\end{equation} 
\noindent Now if we take the general case that the particle is initially at the left to the sink $z_0 < z_s $, 
\begin{equation}
k_I^{-1}= \frac{e^{{z_s}^2/2}}{\frac{ 2 \epsilon k_{0}\sqrt{{\alpha_G}}}{\sqrt{\pi} Erf(\sqrt{{\alpha_G}} \epsilon)}{[B/{(2\pi D)}]}^{1/2} }+ \left[\frac{\partial}{\partial k_r}\left[\frac{e^{[(z_0^2-z_s^2)/4]}D_v{(-z_0)}}{D_v{(-z_s)}}\right]\right]_{v=0}.
\end{equation} 
\noindent After further simplification we get
\begin{equation}
k_I^{-1}= \frac{e^{{z_s}^2/2}}{\frac{ 2 \epsilon k_{0}\sqrt{{\alpha_G}}}{\sqrt{\pi} Erf(\sqrt{{\alpha_G}} \epsilon)}{[B/{(2\pi D)}]}^{1/2}}+ \left(\int_{z_0}^{z_s} dz e^{(z^2/2)}\left[1+erf(z/\sqrt{2})\right]\right)(\pi/2)^{1/2}B.
\end {equation}
\noindent The long-term rate constant $k_L$ is determined by the value of $s$, at which $s + \frac{ 2 \epsilon k_{0}\sqrt{{\alpha_G}}}{\sqrt{\pi} Erf(\sqrt{{\alpha_G}} \epsilon)} F(z_s,s|z_s) \sqrt{\frac{\pi}{{\alpha_G}}} = 0 $. This equation may be expressed as an equation for $\nu (= -s/B)$
\begin{equation}
\nu = D_\nu(-z_c)D_\nu(z_c)\Gamma(1-\nu){\frac{k_0}{Erf(\sqrt{{\alpha_G}} \epsilon)}}{{[B/{(2\pi D)}]}^{1/2}}.
\end{equation}
\noindent For integer values of $\nu$, $D_\nu(z)=2^{-\nu/2}e^{-z^2/4}H_{\nu}(z/\sqrt{2})$, $H_{\nu}$ are Hermite polynomials. $\Gamma(1-\nu)$ has poles at $\nu = 1,2, . . . .$. Our interest is in $\nu \in [0, 1]$, as $k_L = 2 \nu$ for $k_r =0$. If $ {\frac{ 2 \epsilon k_{0}\sqrt{{\alpha_G}}}{\sqrt{\pi} Erf(\sqrt{{\alpha_G}} \epsilon)}}{{[B/{(2\pi D)}]}^{1/2}}\ll 1$, or $z_c \gg 1$ then $\nu \ll 1$ and one can arrive
\begin{equation}
\nu = D_0(-z_s)D_0(z_s){\frac{ 2 \epsilon k_{0}\sqrt{{\alpha_G}}}{\sqrt{\pi} Erf(\sqrt{{\alpha_G}} \epsilon)}}{{[B/{(2\pi D)}]}^{1/2}}, 
\end{equation}
and hence 
\begin{equation}
k_L = {\frac{ 2 \epsilon k_{0}\sqrt{{\alpha_G}}}{\sqrt{\pi} Erf(\sqrt{{\alpha_G}} \epsilon)} e^{{-z_s}^2/2}}{[B/{(2\pi D)}]}^{1/2}.
\end{equation}
\noindent We can see from Eq. (28) and Eq. (31) that the rate constants are depending on the shape of the sink. For higher value of ${\alpha_G}$, Gaussian function will behave like Dirac delta. Now, let us take the ratio of the long time rate constant of Dirac delta sink ${(k_L)_{DD}}$ \cite{Kls1} and narrow Gaussian sink ${(k_L)_{NG}}$, so that
\begin{equation}
\frac{(k_L)_{DD}}{(k_L)_{NG}} = \frac{\sqrt{\pi} Erf(\sqrt{{\alpha_G}} \epsilon)}{2 \epsilon \sqrt{{\alpha_G}}},
\end{equation}
\noindent where ${\alpha_G}$ and $\epsilon$ are always positive so $\frac{\sqrt{\pi} Erf(\sqrt{{\alpha_G}} \epsilon)}{2 \epsilon \sqrt{{\alpha_G}}}$ will very from $1$ to $0$. For smaller values of $\sqrt{\alpha_G} \epsilon$, $\frac{\sqrt{\pi} Erf(\sqrt{{\alpha_G}} \epsilon)}{2 \epsilon \sqrt{{\alpha_G}}}\rightarrow 1$, therefore $\frac{(k_L)_{DD}}{(k_L)_{UG}}\rightarrow 1$ . For larger values of $\sqrt{\alpha_G} \epsilon$, $\frac{\sqrt{\pi} Erf(\sqrt{{\alpha_G}} \epsilon)}{2 \epsilon \sqrt{{\alpha_G}}}\rightarrow 0$, therefore $\frac{(k_L)_{DD}}{(k_L)_{UG}}\rightarrow 0$ . So the $k_L$ for ultra-short Gaussian sink model will be larger then that of Dirac delta function sink model. This can be easily understood by the fact that wider sink provide a lot of reaction channels and therefor effectively accelerate the reaction.

\subsection{Ultra-Short Exponential Sink}

\noindent In this section, we give an exact solution of the problem for a sink, which is represented by a Exponential function of
narrow width.   Exponential sink can be used to model decay processes with one of the example as electron transfer in a single protein \cite{Min}. Now in Eq. (6), we replace $S(x)$ by a truncated Exponential function is given below
\begin{equation}
S(x) =\frac{{\alpha_E}}{2 (1 - e^{ - {\alpha_E} \epsilon})} e^{- {\alpha_E} |x - x_c|} f(x).    
\end{equation}    
\noindent In the above $S(x)$ is a non zero constant within a narrow range of position coordinate $x$ {\it i.e.,} from  $x_c - \epsilon$ to $x_c + \epsilon$, where $\epsilon$ is a very small but positive number. Outside this range the value of $S(x)$ is assumed to be zero. The above sink function is normalized {\it i.e.}, $\int_{-\infty}^{\infty} S(x) dx = 1$. The function $f(x)$ is introduced in the definition of $S(x)$ to ensure that $S(x)$ is non-zero only within $x_c - \epsilon$ and $x_c + \epsilon$ (therefore $f(x)$ is taken to be $1$ within this range and it is zero otherwise). So in Eq. (6) the term $k_{0} S(x){\tilde P}(x,s)$ can be replaced by $k_{0} e^{- {\alpha_E} |x - x_c|}{\tilde P}(x_c,s) f(x)$. So Eq. (6) becomes
\begin{equation}
\left[s {\tilde P}(x,s)-D\frac{\partial^2{\tilde P}(x,s)}{\partial x^2} - \frac{D B}{k_{b}T}\frac{\partial{\tilde P}(x,s)}{\partial x} x + k_{0}\frac{{\alpha_E}}{2 (1 - e^{ - {\alpha_E} \epsilon})}e^{- {\alpha_E} |x - x_c|} {\tilde P}(x_c,s) f(x)+k_{r}{\tilde P}(x,s)\right] =  P(x,0).
\end{equation}
\noindent The solution of the above equation in terms of Green's function $G(x,s|x_0)$ is given below
\begin{equation}
\tilde P(x,s)= \int^\infty_{-\infty} dx_{0}G(x,s+k_{r}|x_0)P(x_0,0) - k_{0} \frac{{\alpha_E}}{2 (1 - e^{ - {\alpha_E} \epsilon})} {\tilde P}(x_c,s)\int^{\infty}_{-\infty} dx_{0}G(x,s+k_{r}|x_0)e^{- {\alpha_E} |x_0 - x_c|} f(x_0).
\end{equation}
\noindent By using the properties of $f(x)$, we can change the limits of integration in the last term of Eq. (35) to get
\begin{equation}
\tilde P(x,s)= \int^\infty_{-\infty} dx_{0}G(x,s+k_{r}|x_0)P(x_0,0) - k_{0} \frac{{\alpha_E}}{2 (1 - e^{ - {\alpha_E} \epsilon})} {\tilde P}(x_c,s)\int^{x_c+\epsilon}_{x_c -\epsilon} dx_{0}G(x,s+k_{r}|x_0) e^{- {\alpha_E} |x_0 - x_c|}.
\end{equation}
\noindent As the range of integration is very small, we safely assume that the relevant Green's function does not change within the range of integration - therefore $G(x,s+k_{r}|x_0)$ in the above equation can be replaced by $G(x,s+k_{r}|x_c)$ to get
\begin{equation}
\tilde P(x,s)= \int^\infty_{-\infty} dx_{0}G(x,s+k_{r}|x_0)P(x_0,0) -\frac{{\alpha_E} k_{0}}{2 (1 - e^{ - {\alpha_E} \epsilon})} {\tilde P}(x_c,s)G(x,s+k_{r}|x_c)\int^{x_c+\epsilon}_{x_c -\epsilon} dx_{0}e^{- {\alpha_E} |x_0 - x_c|}.
\end{equation}
\noindent Now we assume that ${\alpha_E}$ is very large, such that $e^{- {\alpha_E} |x_0 - x_c|}$ is practically zero, outside the range  $x_c - \epsilon$ to $x_c + \epsilon$. Therefore, we can change the limits of the integration over the Exponential function from `$x_c - \epsilon$ to $x_c + \epsilon$' to `$-\infty$ to $\infty$' as shown below
\begin{equation}
\tilde P(x,s)= \int^\infty_{-\infty} dx_{0}G(x,s+k_{r}|x_0)P(x_0,0) - \frac{{\alpha_E} k_{0}}{2 (1 - e^{ - {\alpha_E} \epsilon})} {\tilde P}(x_c,s)G(x,s+k_{r}|x_c)\int^{\infty}_{-\infty} dx_{0}e^{- {\alpha_E} |x_0 - x_c|}.
\end{equation}
\noindent Now the Second term on the R.H.S. of the above equation can be integrated over $x_0$ easily to get
\begin{equation}
\tilde P(x,s)= \int^\infty_{-\infty} dx_{0}G(x,s+k_{r}|x_0)P(x_0,0) -  \frac{k_0 \epsilon \alpha_E}{(1-e^{-{\alpha_E} \epsilon})} {\tilde P}(x_c,s)G(x,s+k_{r}|x_c).
\end{equation}
\noindent  Now the above equation has two unknowns, $\tilde P(x,s)$ and $\tilde P(x_c,s)$, so we put $x=x_c$ and get the following equation with only one unknown {\it i.e.}, $\tilde P(x_c,s)$.
\begin{equation}
\tilde P(x_c,s)= \int^\infty_{-\infty} dx_{0}G(x_c,s+k_{r}|x_0)P(x_0,0) -  \frac{k_0 \epsilon \alpha_E}{(1-e^{-{\alpha_E} \epsilon})} {\tilde P}(x_c,s)G(x_c,s+k_{r}|x_c).
\end{equation}
\noindent Now we solve the above equation for $\tilde P(x_c,s)$ to get
\begin{equation}
\tilde P(x_c,s)= \frac{\int^\infty_{-\infty} dx_{0}G(x_c,s+k_{r}|x_0)P(x_0,0)}{1+\frac{k_0 \epsilon \alpha_E}{(1-e^{-{\alpha_E} \epsilon})}G(x_c,s+k_{r}|x_c)}.
\end{equation}
\noindent This solution, when we substituted back into Eq. (39), we get
\begin{equation}
\tilde P(x,s)= \int^\infty_{-\infty} dx_{0} \left[G(x,s+k_{r}|x_0)-\frac{\frac{k_0 \epsilon \alpha_E}{(1-e^{-{\alpha_E} \epsilon})} G(x,s+k_{r}|x_c)G(x_c,s+k_{r}|x_0)}{1+ \frac{k_0 \epsilon \alpha_E}{(1-e^{-{\alpha_E} \epsilon})} G(x_c,s+k_{r}|x_c)}\right] P(x_0,0).
\end{equation}
\noindent Survival probability in Laplace domain can be easily calculated by using the following equation
\begin{equation}
P_e(s) = \int^\infty_{-\infty} dx {\tilde P}(x,s).
\end{equation} 
\noindent  We get the following expression of survival probability, using Eq. (42) and Eq. (43) - which is given by
\begin{equation}
P_e(s)=\frac{1}{s+k_{r}}\left[1-\left(1+ \frac{k_0 \epsilon \alpha_E}{(1-e^{-{\alpha_E} \epsilon})} G(x_c,s+ k_{r}|x_c) \right)^{-1} \frac{k_0 \epsilon \alpha_E}{(1-e^{-{\alpha_E} \epsilon})} \int^\infty_{-\infty} dx_0 G (x_c,s+k_{r}|x_0)P(x_0,0)\right].
\end{equation}
\noindent The average rate constant of the reaction can be derived using Eq. (44) and is given by  
\begin{equation}
{k_I}^{-1}= \frac{1}{k_{r}}\left[1-\left(1+ \frac{k_0 \epsilon \alpha_E}{(1-e^{-{\alpha_E} \epsilon})} G(x_c, k_{r}|x_c) \right)^{-1} \frac{k_0 \epsilon \alpha_E}{(1-e^{-{\alpha_E} \epsilon})} \int^\infty_{-\infty} dx_0 G (x_c,k_{r}|x_0)P(x_0,0)\right].
\end{equation}
\noindent If we assume the initial condition $P^0_e(x_0)$ is represented by a Dirac delta function located at $x_0$. Then the analytical expression of average rate constant is given by
\begin{equation}
{k_I}^{-1}= \frac{1}{k_{r}}\left[1-\left(1+ \frac{k_0 \epsilon \alpha_E}{(1-e^{-{\alpha_E} \epsilon})} G(x_c, k_{r}|x_c) \right)^{-1} \frac{k_0 \epsilon \alpha_E}{(1-e^{-{\alpha_E} \epsilon})} \int^\infty_{-\infty} dx_0 G (x_c,k_{r}|x_0)P(x_0,0)\right].
\end{equation}
\noindent The analytical expression for long term rate constant is given by
\begin{equation}
k_L= k_r - [ values \; of \; s \; for \; which \;\; s+ {\frac{k_0 \epsilon \alpha_E}{(1-e^{-\alpha_E\epsilon})}} F(z_s,s|z_s)=0].
\end{equation}
\noindent Using the properties of $D_v{(z)}$, we find that when $k_r\rightarrow 0, F{(z_s,k_r|z_0)}$ and $F{(z_s,k_r|z_s)}\rightarrow exp(-z_s^2/2){{[B/{(2\pi D)}]}^{1/2}}$ so that
\begin{equation}
  \frac{k_0 \epsilon \alpha_E}{(1-e^{-{\alpha_E} \epsilon})} F{(z_s,k_r|z_0)}/[k_r+   \frac{k_0}{(1-e^{-{\alpha_E} \epsilon})} F{(z_s,k_r|z_s)}]\rightarrow 1.
\end{equation}
\noindent Hence, we get the following expression for $k_{I}$
\begin{equation}
k_I^{-1}=-{\frac{\partial}{\partial k_r}\left[\frac{\frac{k_0 \epsilon \alpha_E}{(1-e^{-{\alpha_E} \epsilon})}  F(z_s,k_r|z_0)}{k_r + \frac{k_0 \epsilon \alpha_E}{(1-e^{-{\alpha_E} \epsilon})} F(z_s,k_r|z_s)}\right]}_{(k_r) \rightarrow 0}.
\end{equation} 
\noindent If the particle is initially at the left of sink, then we take $z_0 < z_s $. Therefore we get the following expression of rate constant
\begin{equation}
k_I^{-1}= \frac{e^{{z_s}^2/2}}{ \frac{k_0 \epsilon \alpha_E}{(1-e^{-{\alpha_E} \epsilon})}{[B/{(2\pi D)}]}^{1/2}}+ \left[\frac{\partial}{\partial k_r}\left[\frac{e^{[(z_0^2-z_s^2)/4]}D_v{(-z_0)}}{D_v{(-z_s)}}\right]\right]_{v=0}.
\end{equation} 
\noindent After simplification we get the following expression
\begin{equation}
k_I^{-1}= \frac{e^{{z_s}^2/2}}{\frac{k_0 \epsilon \alpha_E}{(1-e^{-{\alpha_E} \epsilon})}{[B/{(2\pi D)}]}^{1/2}}+ \left(\int_{z_0}^{z_s} dz e^{(z^2/2)}\left[1+erf(z/\sqrt{2}\right]\right){(\pi/2)}^{1/2} B.
\end {equation}
\noindent The long-term rate constant $k_L$ is determined by the value of $s$, which satisfy $s+  \frac{k_0 \epsilon \alpha_E}{(1-e^{-{\alpha_E} \epsilon})} F(z_s,s|z_s)  =0 $. This equation may be written as an equation for $\nu (= -s /B)$
\begin{equation}
\nu = D_\nu(-z_c)D_\nu(z_c)\Gamma(1-\nu){\frac{k_0 \epsilon \alpha_E}{(1-e^{-{\alpha_E} \epsilon})}}{{[B/{(2\pi D)}]}^{1/2}}.
\end{equation}
\noindent Taking the integer values of $\nu$, $D_{\nu} (z) =2^{-\nu/2}e^{-z^2/4}H_{\nu}(z/\sqrt{2})$, where $H_{\nu}$ are Hermite polynomials. $\Gamma(1-\nu)$ have poles at $\nu = 1,2, . . . .$. Our interest is in the values of $\nu$ in the range $\nu$ in [0, 1], as $k_{L} = 2 \nu$ for 
$k_r =0$. If $ {\frac{k_0 \epsilon \alpha_E}{(1-e^{-{\alpha_E} \epsilon})}}{{[B/{(2\pi D)}]}^{1/2}}\ll 1$, or $z_{s} \gg 1$ then $\nu \ll 1$ and one can arrive
\begin{equation}
\nu = D_0(-z_s)D_0(z_s){\frac{k_0 \epsilon \alpha_E}{(1-e^{-{\alpha_E} \epsilon})} }{{[B/{(2\pi D)}]}^{1/2}}. 
\end{equation}
\noindent The analytical expression for long-time rate constant is given below 
\begin{equation}
k_L = {\frac{k_0 \epsilon \alpha_E}{(1-e^{-{\alpha_E} \epsilon})} e^{{-z_s}^2/2}}{[B/{(2\pi D)}]}^{1/2}.
\end{equation}
\noindent  We can see from Eq. (51) and Eq. (54) that the rate constants are depending on the shape of the sink. Now, let us take the ratio of the long time rate constant for Dirac delta sink model ${(k_L)_{DD}}$ \cite{Kls1} and ultra-short Exponential sink model ${(k_L)_{UE}}$, so that we get
\begin{equation}
\frac{(k_L)_{DD}}{(k_L)_{UE}} = \frac{(1-e^{-{\alpha_E} \epsilon})}{\epsilon \alpha_E}.
\end{equation}
\noindent where ${\alpha_E}$ and $\epsilon$ are always positive so $\frac{(1-e^{-{\alpha_E} \epsilon})}{\epsilon \alpha_E}$ will very from $1$ to $0$. Therefore, $\frac{(k_L)_{DD}}{(k_L)_{UE}}$ will vary between $1$ and $0$. 
For small values of ${\epsilon \alpha_E}$,  $\frac{(k_L)_{DD}}{(k_L)_{UE}}\rightarrow 1$ and for larger values of ${\epsilon \alpha_E}$,  $\frac{(k_L)_{DD}}{(k_L)_{UE}}\rightarrow  0$. Therefore the $k_L$ for ultrashort Exponential sink model will be larger then that of Dirac delta function sink model. This can be easily understood by the fact that wider sinks provide more reaction channels and thus effectively accelerate the reaction.  Now, let us take the ratio of the long time rate constant of ultra-short Gaussian sink ${(k_L)_{UG}}$  and ultra-short Exponential sink ${(k_L)_{UE}}$, so
\begin{equation}
\frac{(k_L)_{UG}}{(k_L)_{UE}} = \frac{(1-e^{-{\alpha_E} \epsilon}) 2 \sqrt{\alpha_G}}{ \sqrt{\pi} Erf(\sqrt{\alpha_G} \epsilon) {\alpha_E}}.
\end{equation}
\noindent  For relatively smaller values of $\epsilon$,  $\frac{(k_L)_{UG}}{(k_L)_{UE}}\rightarrow 1$ and for relatively larger values of $\epsilon$,  $\frac{(k_L)_{NG}}{(k_L)_{NE}}$ different from $1$. So the $k_L$ for ultra-short Exponential sink model will be different from that of ultra-short Gaussian sink model.

\subsection{Ultra-short Rectangular sink}
\noindent In this section, we give an exact solution of the problem for a sink, which is represented by a non-zero constant of
narrow width. Exponential sink can be understood as the most simplest generalization od point sink model. Therefore in Eq. (6), we replace $S(x)$ by a truncated rectangular function as given below
\begin{equation}
S(x) =\frac{ f(x)}{2 \epsilon}.    
\end{equation}    
\noindent In the above, the value of $S(x)$ is $1$ in a narrow range of position coordinate $x$ {\it i.e.,} from  $x_c - \epsilon$ to $x_c + \epsilon$, where $\epsilon$ is a very small positive number. Outside this range the value of $S(x)$ is zero. The above sink function is normalized {\it i.e.}, $\int_{-\infty}^{\infty} S(x) dx = 1$. The function $f(x)$ is introduced in the above equation just to ensure that $S(x)$ is $1$ only within $x_c - \epsilon$ and $x_c + \epsilon$ (therefore $f(x)$ is assumed to be $1$ within this range and it is zero otherwise). So in Eq.(6) the term $k_{o} S(x){\tilde P}(x,s)$ can be replaced by $k_{0}{\tilde P}(x_c,s) f(x)$. So Eq. (6) becomes
\begin{equation}
\left[s {\tilde P}(x,s)-D\frac{\partial^2{\tilde P}(x,s)}{\partial x^2} - \frac{D B}{k_{b}T}\frac{\partial{\tilde P}(x,s)}{\partial x} x + \frac{k_{0}}{2 \epsilon} {\tilde P}(x_c,s) f(x)+k_{r}{\tilde P}(x,s)\right] =  P(x,0).
\end{equation}
\noindent The solution of the above equation in terms of Green's function $G(x,s|x_0)$ is given below
\begin{equation}
\tilde P(x,s)= \int^\infty_{-\infty} dx_{0}G(x,s+k_{r}|x_0)P(x_0,0) - \frac{k_{0}}{2 \epsilon}{\tilde P}(x_c,s)\int^{\infty}_{-\infty} dx_{0}G(x,s+k_{r}|x_0)f(x_0).
\end{equation}
\noindent By using the properties of $f(x)$, Eq. (59) can be further simplified to
\begin{equation}
\tilde P(x,s)= \int^\infty_{-\infty} dx_{0}G(x,s+k_{r}|x_0)P(x_0,0) - \frac{k_{0}}{2 \epsilon} {\tilde P}(x_c,s)\int^{x_c+\epsilon}_{x_c -\epsilon} dx_{0}G(x,s+k_{r}|x_0).
\end{equation}
\noindent  As the range of integration is very small, we safely assume that the relevant Green's function does not change within the range of integration - therefore $G(x,s+k_{r}|x_0)$ in the above equation can be replaced by $G(x,s+k_{r}|x_c)$ to get
\begin{equation}
\tilde P(x,s)= \int^\infty_{-\infty} dx_{0}G(x,s+k_{r}|x_0)P(x_0,0) - \frac{k_{0}}{2 \epsilon} {\tilde P}(x_c,s)G(x,s+k_{r}|x_c)\int^{x_c+\epsilon}_{x_c -\epsilon} dx_{0}.
\end{equation}
\noindent After performing the integration over $x_0$, in the second term on the R.H.S. of the above equation we get
\begin{equation}
\tilde P(x,s)= \int^\infty_{-\infty} dx_{0}G(x,s+k_{r}|x_0)P(x_0,0) - k_{0} {\tilde P}(x_c,s)G(x,s+k_{r}|x_c).
\end{equation}
\noindent  Now the above equation has two unknowns $\tilde P(x,s)$ and $\tilde P(x_c,s)$, so we put $x=x_c$ and get the following equation with one unknown {\it i.e.}, $\tilde P(x_c,s)$.
\begin{equation}
\tilde P(x_c,s)= \int^\infty_{-\infty} dx_{0}G(x_c,s+k_{r}|x_0)P(x_0,0) -  k_0  {\tilde P}(x_c,s)G(x_c,s+k_{r}|x_c).
\end{equation}
\noindent Now we solve the above equation for $\tilde P(x_c,s)$ to get
\begin{equation}
\tilde P(x_c,s)= \frac{\int^\infty_{-\infty} dx_{0}G(x_c,s+k_{r}|x_0)P(x_0,0)}{1+k_0 G(x_c,s+k_{r}|x_c)}.
\end{equation}
\noindent This solution, when we substituted back into Eq. (62) we get
\begin{equation}
\tilde P(x,s)= \int^\infty_{-\infty} dx_{0} \left[G(x,s+k_{r}|x_0)-\frac{ k_0 G(x,s+k_{r}|x_c)G(x_c,s+k_{r}|x_0)}{1+ k_0 G(x_c,s+k_{r}|x_c)}\right] P(x_0,0).
\end{equation}
\noindent The above equation is the same equation, that one gets in case of Dirac delta function sink model \cite{Kls1}. For this case time averaged rate constant is given by \cite{Kls1}
\begin{equation}
{k_I}^{-1}= (k_r)^{-1}\left(1 - \frac{ k_0 F(z_s,k_r|z_0)}{k_r+ k_0 F(z_s,k_r|z_s)} \right),
\end{equation}
\noindent and the long time rate constant is given by
\begin{equation}
k_L = k_0 e^{-z_s^2/2}[B/{(2\pi D)}]^{1/2}.
\end{equation}

   
\section{Conclusions}
 \noindent We have given a method for solving the problem of radiationless decay, modelled by a modified Smoluchowski equation for a harmonic potential with a sink of ultrashort width. In one case, the sink is represented by a narrow Gaussian function, in another case the sink is represented by a narrow Exponential function. Also we have considered the case rectangular shape. In all cases, the exact analytical expressions for $k_{L}$ and $k_{I}$ are derived. Both $k_{L}$ and $k_{I}$ are found to be dependent on the shape of the sink function as well as on the width of the sink $\epsilon$. These new models has been compared with the earlier Dirac delta sink model. Our method can easily be extended to the cases where sink is represented by a collections of narrow Gaussian or Exponential functions.

\section{Acknowledgments:}
\noindent One of the author (S.M.) would like to thank IIT Mandi for HTRA fellowship and the other author (A.C.) thanks IIT Mandi for providing PDA grant.


\section {References}
\begin{thebibliography}{99}

\bibitem{Harris} D. Ben-Amotz and C. B. Harris, J. Chem. Phys. {\bf 86} (1987) {\it5433}.

\bibitem{Ben} D. Ben-Amotz and C. B. Harris, J. Chem. Phys. {\bf 86} (1987) {\it 4856}.

\bibitem{Kls1} K. L. Sebastian, Phys. Rev. A  {\bf 46}, {\it R1732}  (1992). 

\bibitem{Ani1} A. Chakraborty, J. Chem. Phys. {\bf 139} (2013) {\it 094101}.

\bibitem{Bagchi} B. Bagchi, J. Chem. Phys. {\bf 87} (1987) {\it 5393}.

\bibitem{lippert} E. Lippert, W. Rettig, V. Bonacic-Koutecky, F. Heisel and J. A. Miehe, Adv. Chem. Phys. {\bf68} (1987) {\it 1}.

\bibitem{SK} A. Samanta and S. K. Ghosh, Phys. Rev. E {\bf 47} (1993) {\it 4568}.

\bibitem{Ani2} A. Chakraborty, arxiv:1308.1354, (2013).

\bibitem{Robin} G. W. Robinson and R. P. Forsch, J. Chem. Phys. {\bf 38} (1963) {\it 1187}.

\bibitem{Amb} A. M. Berezhkovskii, Yu. D. D'yakov and V. Yu. Zitserman, J. Chem. Phys. {\bf 109} (1998) {\it 4182}.

\bibitem{Kls3} N. Chakravarti, K. L. Sebastian, Chem. Phys. Lett., {\bf 204} (1993) {\it 496}.

\bibitem{SSZ} K. Schulten, Z. Schulten and A. Szabo, Physica A, {\bf 100} (1980) {\it 599}.

\bibitem{Oster} G. Oster and N. NishiJama, J. Am. Chem. Soc., {\bf 78} (1956) {\it 1581}.

\bibitem{Fleming} B. Bagchi, G. R. Fleming, J. Phys. Chem.,{\bf94}, (1990), {\it 9}.

\bibitem{Kls2}  K. L. Sebastian, Chem. Sci. , {\bf 106} (1994) {\it 493}.

\bibitem{SM} S. Mudra, H. Chhabra and A. Chakraborty (submitted).

\bibitem{Hansen} Y. Hansen, R.R. Netz, and M. Hinczewski, J. Chem. Phys. {\bf 132}, (2010),{\it135103}.

\bibitem{Min} W. Min, G. Luo, B.J. Cherayil, S.C. Kou, X.S. Xie, Phys. Rev. Lett. {\bf94} (2005){\it 198302}.\cite{Min}

\bibitem{bagchi1} B. Bagchi, G. R. Fleming and D. W. Oxtoby, J. Chem. Phys. {\bf 78} (1983) {\it7375}.

\bibitem{Alok} A. Samanta and S. K. Ghosh, Phys. Rev. E {\bf 47} (1993) {\it 4568}.

\bibitem{Erdelyi} Higher Transcendental Functions, edited by A. Erdelyi(McGraw-Hill, New York, 1953), Vol. II, p. 115

\bibitem{Hilbert} R. Courant and D. Hilbert, Methods of Mathematical Physics Physics, Vol:1; p.351, (Wiley Eastern, 1975).


\end{thebibliography}
\end{document}